# Theoretical study of the stripline ferromagnetic resonance response of metallic ferromagnetic films based on an analytical model

R.Hai[1,2] and M. Kostylev[1]

1. School of Physics, University of Western Australia, Crawley, W.A. 6009, Australia

2. School of Physics, Nanjing University, Nanjing 210093, China

*Abstract:* We develop an advanced analytical model for calculating the broadband stripline ferromagnetic resonance (FMR) response for metallic ferromagnetic films, taking into account the exchange interaction as well as the exchange boundary conditions at the film surface. This approach leads to simple analytical expressions in the Fourier space. As a result, a numerical code which implements inverse Fourier transform of these equations is very quick. This allows us to explore a wide space of parameters as numerical examples of application of this theory. In particular, we investigate the joint effect of microwave eddy current shielding and magnetisation pinning at the ferromagnetic film surfaces on the shape of the stripline FMR response of the film.

## I. INTRODUCTION

Stripline broadband ferromagnetic resonance (FMR) response has been receiving a significant attention in recent years because of its potential for characterisation of magnetic thin films and nanostructures [1-3]. Also, the stripline geometry is important for various applications of magnetization dynamics, such as sensing fields, particles and substances [4-6] and in microwave spintronics [7].

The geometry of a stripline ferromagnetic resonance experiment setup usually contains a microscopic coplanar (CPW) or microstrip line (MSL) through which a microwave current flows (direction $z$ in Fig. 1). In this work we will be focusing on MSL, and Fig. 1 reflects this geometry. A ferromagnetic film which is to be characterized, sits on top of the microstrip line. We will be dealing with metallic films; therefore an insulating spacer separates the film from the microstrip in order to avoid an electrical contact between them. A static magnetic field **H** is applied along the microstrip line. Precession of magnetization in the material is driven by the Oersted field of a microwave current flowing through the microstrip line. On resonance, the amplitude of precession increases sharply which is seen as an increase in the microstrip line transmission losses or a decrease in the microstrip line transmission coefficient S21 [1].

A number of important peculiarities of the stripline FMR response have been recently discovered. For instance, it has been shown that the linewidth of the FMR peaks for the samples as measured with this method can be broadened by excitation of travelling spin waves [8] and over-coupling to the probing stripline ("radiation losses") [9].

Also, it has been demonstrated both experimentally [10-13] and theoretically [14-16] that the geometry of Fig. 1 breaks the symmetry of the microwave magnetic field incidence on the sample surface. This leads to excitation of microwave eddy currents in a sample under test if it is conducting. Consequences of this are especially important for samples with high conductivity of metals - see Fig. 21 in [1] for details of the microwave magnetic field configuration in this case.

Ultimately, the eddy currents lead to strong excitation of standing spin wave modes (SSWM) for those ferromagnetic-film samples, for which one would expect much smaller or even vanishing SSWM amplitudes in the conditions of the conventional cavity FMR [10-11].

The first theoretical paper on this subject treated the stripline FMR response only in the limit of a very large microstrip width [15]. Later on, it was found that the width of the microstrip line has a large effect on the response [16,17]. In particular, it was shown that the decrease in the microstrip width decreases the FMR peak amplitudes corresponding to excitation of the standing spin wave modes [16].

However, the theoretical approaches from those publications have significant drawbacks which did not allow exploring the impact of the finite width of the stripline on the entire space of parameters for the stripline FMR experiment. The numerical model constructed in [16] is just very slow. The analytical solution from [17] delivers quantitative results almost instantly; however, it was obtained in the exchange-free approximation, hence it allows simulation of only the fundamental FMR peak.

In this work we fill the gap and report an analytical solution for the dipole-exchange case and carry out a number of calculations by using the derived model. The obtained results agree well with the fully numerical model from [16]. Furthermore, since a numerical code we developed based on this analytical solution is very fast, we are now able to explore a range of details of the broadband FMR experiment which have not been studied before. These are the effects of the finite width of the stripline and of the spacer thickness on the amplitudes of the response of the standing-spin wave modes. Also, we are now able to consider radiation losses for the standing wave excitations and the effect of surface anisotropies (surface spin pinning) on these peaks.

The paper is organized as follows. In Section II we construct the analytical model. In Section III it is used to obtain a number of quantitative results which are discussed in detail in that section.    Section IV contains conclusions.

II. Theory

We consider a metallic ferromagnetic film of thickness $L$ which is homogeneous throughout its volume, but may possess perpendicular surface anisotropy at one or both surfaces. In the model, the perpendicular anisotropy is introduced as surface magnetization pinning [18]. We treat the film as infinitely long along both in-plane axes - $x$ and $z$. The external magnetic field $H$ is applied in the positive $z$ direction. The film sits on top of a dielectric spacer of a thickness $s$ separating the film from a microstrip of a width $w$. We neglect the fact that real microstrips have a small (with respect to $w$), but finite thickness and treat the microstrip as infinitely thin in the direction y. The microstrip is supported by a dielectric substrate of thickness $d$ whose second surface is uniformly metalized. Although the magnetostatic approximation utilized below does not need specifying the dielectric constant for the substrate, we implicitly assume a specific value for it. It is one which results in a 50 Ω of characteristic impedance for the microstrip line in the absence of a film on its top.

One standard approach to calculation of the complex impedance for the microstrip lines loaded by ferromagnetic films is by exploiting the translation invariance in the $z$-direction [19]. This corresponds to a quasi-static approach for description of microwave transmission lines and results in a two-dimensional problem which we will be dealing with below. The actual form of the electromagnetic field dependence on the $z$ co-ordinate – in the form of an electromagnetic wave travelling along the stripline - will be accounted for while employing the obtained expression for the complex impedance for calculations of the transmission coefficient S21.

In order to describe the magnetization dynamics in the film we employ the linearized Landau-Lifschitz equation (LLE)

$$\frac{\partial \mathbf{m}}{\partial t} = -|\gamma|(\mathbf{m} \times \mathbf{H} + \mathbf{M} \times \mathbf{h}_{eff}) , \qquad (1)$$

where $\mathbf{H} = H_0 \mathbf{u}_z$, $\mathbf{M} = M_0 \mathbf{u}_z$, and $\mathbf{u}_z$ is a unit vector in the direction z. In Eq.(1), the dynamic magnetization $m$ has only two components: $\mathbf{m} = (m_x, m_y)$. They are perpendicular to the static magnetization $\mathbf{M}$ whose magnitude $M_0$ is equal to the saturation magnetization of the film $|M_0| = M_S$. The dynamic effective field $\mathbf{h}_{eff}$ consists of two parts: the effective exchange field $\mathbf{h}_{ex}$ is given by

$$\mathbf{h}_{ex} = \alpha \left( \frac{\partial^2}{\partial x^2} + \frac{\partial^2}{\partial y^2} \right) \mathbf{m} , \qquad (2)$$

and the dynamic magnetic field $h$ being solution of Maxwell equations in the electric-bias free approximation

$$\nabla \times \mathbf{h} = \sigma \mathbf{e} , \qquad (3)$$
$$\nabla \cdot \mathbf{h} = -\nabla \cdot \mathbf{m} , \qquad (4)$$
$$\nabla \times \mathbf{e} = -i\omega\mu_0 (\mathbf{h} + \mathbf{m}) . \qquad (5)$$

Here $e$ is the microwave electric field, $\sigma$ is electric conductivity, $\mu_0$ is the permittivity of vacuum and $\omega$ is the frequency of the microwave driving field. Dynamic magnetic field $h$ has two components $\mathbf{h} = (h_x, h_y)$ which are perpendicular to the direction along the stripline. Note that the term containing electric permittivity is missing on the right hand side of Eq. (3); this is consistent with the standard magnetostatic approximation for ferromagnetic materials [20]. Furthermore, for metals at microwave frequencies, the contribution of the electric bias field to Eq.(3) is negligible with respect to the conductivity one [21,15,17].

Even in two dimensions, previous calculations faced serious difficulties when trying to employ real-space methods [1,14] because of the incompatibility of the length scales for L, w and d. By exploiting the translational symmetry in the x-direction, that is, by applying the spatial Fourier transformation to both sides of Eqs. (1-5) we can significantly simplify the problem. This is because the microstrip has a infinitely small thickness in the y-direction and hence can be treated as a boundary condition involving the surface current density [19,22,16,17]. The surface current and its distribution along x are assumed to be given. Alternatively, it can be calculated self-consistently [17,19], but the latter is out of scope of the present paper.

One more important advantage of the Fourier-space approach is that simple analytical solutions exist for the areas above and below the film [16,17]. This allows one to exclude those areas from consideration and consider only the dynamic equations for the film. Exclusion of those areas produces specific boundary conditions for the electromagnetic fields on the film surfaces [17].

According to the Fourier space approach, we have

$$\mathbf{m}, \mathbf{h} \sim \exp(i\omega t - ikx) . \qquad (6)$$

Substituting Eq. (6) into Eqs. (3-5), we have

$$k e_z = -\omega\mu_0 (h_y + m_y) , \qquad (7)$$

$$\frac{\partial h_x}{\partial y} + ikh_y = -\sigma e_z, \tag{8}$$

$$-ikh_x + \frac{\partial h_y}{\partial y} = -\frac{\partial m_y}{\partial y} + ikm_x, \tag{9}$$

$$\frac{\partial e_z}{\partial y} = -i\omega\mu_0(h_x + m_x). \tag{10}$$

It is easy to verify that, similar to Ref. [21] the ansatz $\mathbf{h}, \mathbf{m} \sim e^{qy}$ solves the system of equations (1), (7-10). Eqs. (7-10) then reduce to

$$kqh_x + iK^2 h_y - i(k^2 - K^2)m_y = 0, \tag{11}$$

$$-ikh_x + qh_y + qm_y - ikm_x = 0, \tag{12}$$

where $K^2 = k^2 + i\omega\mu_0$,

and Eq.(1) takes the form as follows:

$$h_x = \left[\frac{\omega_H}{\omega_M} + \alpha(k^2 - q^2)\right] m_x - i\frac{\omega}{\omega_M} m_y, \tag{14}$$

$$h_y = \left[\frac{\omega_H}{\omega_M} + \alpha(k^2 - q^2)\right] m_y + i\frac{\omega}{\omega_M} m_x, \tag{15}$$

where $\omega_H = \gamma(H + i\Delta H)$, $\omega_M = \gamma M_S$, and $\Delta H = \alpha_G \omega / \gamma$ is the magnetic loss parameter which scales as the Gilbert magnetic damping constant $\alpha_G$. The magnetic losses have been introduced into Eqs. (14,15) phenomenologically, as it is known that the linearized Landau-Lifshitz-Gilbert Equation reduces to the linearised Landau-Lifshitz Equation (1) with a complex value of the applied field given by the expression above [20]. Equations (11-15) form a homogeneous system of linear algebraic equations. On elimination of the $h_x$ and $h_y$ variables, the system reduces to two equations (A1) for $m_x$ and $m_y$ shown in Appendix.

Equating the determinant of the system (A1) to zero produces a characteristic equation for the problem. The above-mentioned approach of first eliminating $h_x$ and $h_y$ from the system has an important advantage – it results in a compact characteristic equation which is bi-cubic with respect to $q$:

$$aQ^3 + bQ^2 + cQ + d_0 = 0, \tag{16}$$

where $Q=q^2$ and the coefficients $a$, $b$, $c$ and $d_0$ are given in Appendix (Eqs.(A2)). The sixth order for the characteristic equation is in agreement with the earlier treatment of the dipole-exchange spectrum [21] and, as we will see below, with the number of available boundary conditions.[1]

The cubic equation (16) allows an analytic solution. For completeness, it is also given in the appendix, although this is a textbook result. The three roots of (16) are $Q_1$, $Q_2$ and $Q_3$ (see Eqs.(A3)).

---

[1] Direct evaluation of the determinant of (11-14) results in an equation of $8^{th}$ order with respect to $q$, but as the method of elimination of the $h_x$ and $h_y$ variables demonstrates, two of eight roots in total of the $8^{th}$ order equation are meaningless.

Accordingly, the 6 roots of the characteristic equation read: $q_1 = \sqrt{Q_1}$, $q_2 = \sqrt{Q_2}$, $q_3 = \sqrt{Q_3}$, $q_4 = -\sqrt{Q_1}$, $q_5 = -\sqrt{Q_2}$, $q_6 = -\sqrt{Q_3}$. Note that since the coefficients of Eq.(16) are complex numbers, the roots are also complex.

The presence of the six roots indicates that the complete solution for the two components of **m** can be expressed in the following form

$$m_y = \sum_{i=1}^{6} M_i^y \exp(q_i y), \quad (17)$$

$$m_x = \sum_{i=1}^{6} M_i^x \exp(q_i y). \quad (18)$$

In order to determine the coefficients $M_x^i$ (or $M_y^i$), we need six boundary conditions at the film surfaces. Furthermore, in order to obtain a nontrivial solution, at least one of these boundary conditions should be inhomogeneous.

The first set of available boundary conditions is the exchange boundary conditions for the original Landau-Lifshitz equation. They apply to the magnetization vector [18]

$$\frac{\partial \mathbf{m}}{\partial y} \pm d_p \mathbf{m} = 0. \quad (19)$$

Where $d_p$ is the surface magnetization pinning constant. The sign in front of $d_p$ is negative for the surface $y = 0$ while it is positive for the surface $y = L$. The pinning constant is non-vanishing if surface anisotropies are present at the film surfaces. For instance, in the case of a perpendicular uni-axial surface anisotropy which will be considered in Section III, the $m_y$ component can be pinned but the $m_x$ component remains free ($d_p^y \neq 0, d_p^x = 0$). In the case of an in-plane uni-directional in-plane anisotropy (exchange bias), both pinning constants are non-vanishing and depend on the direction of the exchange bias field with respect to **H** [23].

As **m** has two components - $m_x$ and $m_y$, and this exchange boundary condition applies to both surfaces of the film and to the two **m**-vector components separately, altogether four independent boundary conditions are available.

Two more boundary conditions are still needed. These are electromagnetic boundary conditions which follow from Maxwell Equations. As shown in [17], for the geometry of Fig. 1, they can be cast in the form which involves field components *inside* the film only. For the surface $y = L$ the respective boundary condition reads

$$\frac{|k|}{k}(h_y + m_y) + i h_x = 0. \quad (20)$$

Here we would like to recall that we seek solution of the problem in the Fourier space, therefore the field and magnetization vector components entering Eq.(20) are actually spatial Fourier components of these quantities.

For the surface $y = 0$ the boundary condition is inhomogeneous because of the presence of a microwave current $I$ flowing through the microstrip in the direction $z$. The linear density of this current $j(x)$ is assumed to be given. Its spatial Fourier transform reads: $j_k = \frac{1}{2\pi} \int_{-\infty}^{\infty} j(x)\exp(ikx)dx$.

The inhomogeneous boundary condition for $y=0$ reads

$$(h_y + m_y)\coth(|k|(d+s)) - i\frac{|k|}{k}h_x = \frac{\sinh(|k|d)}{\sinh(|k|(d+s))} i\frac{|k|}{k} j_k. \tag{21}$$

Substitution of the solution (17-18) together with the relations between the components of **m** and **h** (Eqs. (A4) in Appendix) into the boundary conditions results in a system of 6 algebraic equations (Eqs.(A5) in Appendix). The system is non-homogeneous, as one of the equations has a non-vanishing right-hand side which follows from the right-hand-side of Eq.(21).

In our work, we solve the system of equations (A5) using numerical methods of linear algebra. The solution is obtained for the $q$-values which solve the characteristic equation (16). As the order of the vector-matrix equation (A6) is just 6, the numerical solution is instantaneous, while run on a personal computer.

The obtained numerical solution for **m** is employed to calculate the linear impedance $Z_r$ of the microstrip loaded by the film, since the latter quantity is an indicator of the microwave magnetic absorption by the film [15]. The linear impedance can be defined as follows:

$$Z_r = \frac{U}{I}, \tag{22}$$

where $U$ is the linear voltage along the microstip which can be defined as the mean value of the total electric field $e_z(x)$ induced at the surface of the strip

$$U = -\frac{1}{w} \int_{-w/2}^{w/2} e_z(x, y = -s)dx. \tag{23}$$

Keeping in mind that our solution is in the Fourier space it is useful to express $U$ in terms of the Fourier components $e_{zk}$ of $e_z(x)$ ($e_{zk} = \frac{1}{2\pi} \int_{-\infty}^{\infty} e_z(x)\exp(ikx)dx$). From the solution of Eqs.(7-10) for the area $y<0$ (characterized by $\sigma=0$, **m**=0) it follows that

$$e_{zk}(y = -s) = -\frac{i\omega\mu_0}{|k|}\sinh(|k|d)\left(\frac{\cosh(|k|d)j_k}{\sinh[|k|(d+s)]} + \frac{h_x(y=0)}{\cosh[|k|(d+s)]}\right), \tag{24}$$

where the Fourier transform of the magnetic field at the film surface $y=0$ reads

$$h_x(y = 0) = \sum_{i=1}^{6} C_{hx}(q_i) M_i^x. \tag{25}$$

Also, it is convenient to express $U$ in terms of the Fourier components of $e_z(x)$. This expression reads

$$U = -\int_{-\infty}^{\infty} e_{zk} \frac{\sin\left(\frac{kw}{2}\right)}{\frac{kw}{2}} dk. \tag{26}$$

This concludes the solution for $Z_r$. Similar to our previous works [16,17], the integral in (26) is computed numerically in the present work.

Thus, the complete process of the numerical simulation consists of three steps. The first one (Step 1) is to calculate the roots $q_1$ to $q_6$ by using the expressions (A3). Because of the analytical character of this solution, the program code implementing it delivers those values instantaneously. The next step is to substitute these $q$-values into Eqs. (A5) and solve the inhomogeneous system of linear algebraic equations numerically. This gives the six amplitudes $M_x^i$. This computation is practically instantaneous, as already stated above. Substitution of the obtained $M_x^i$ values into Eq.(25) and the result into Eq.(24) concludes Step 2. Thus, the output of Step 2 is a value of $e_{zk}$ for some given value of $k$.

The two steps have to be repeated numerous times, as we need $e_{zk}$ values for a large number of Fourier wave numbers $k$, in order to implement Step 3 – carrying out the numerical integration in Eq.(26).

In our previous work [16] up to ten hours of computation were needed to produce the final result - a $Z_r$ dependence on the applied field $H$ for a given microwave frequency. The computation was slow because the model developed in [16] was based on a numerical solution for Step 1. Half a minute or so was needed to complete this step. As it is repeated numerous times for different $k$ and $H$ values, the total computation time becomes very large.

On the contrary, the present code is very fast. It has been implemented as MatLAB and MathCAD worksheets for a usual personal computer. In both cases a complete $Z_r(H)$ dependence is obtained within 10 to 20 seconds. It takes a couple of more seconds to convert the obtained $Z_r$ values into S21 by using Eqs. (18-25) from [16].

## III. DISCUSSION

Several numerical tests were run to validate our theory. First we checked the value of Im($Z_r$) off the resonance peaks (e.g. for $H=0$ or a very large $H$) and for $\sigma=0$ (a non-conducting film). We found that Im($Z_r(H=0, \sigma=0)$) is in excellent agreement with an analytical formula for the linear inductive impedance for a microstrip line not loaded by a ferromagnetic film (see [16,17] for detail).

We also checked the off-resonance distribution of $h_x$ across the film thickness for a highly conducting film. For a 50nm-thick film, $w=1.5$mm and $s=0$, we found that $h_x(x=0, y=0) = j(x=0)$ and $h_x(x=0, y=L)$ is practically vanishing, as expected. Furthermore, the $h_x(y)$ dependence is very close to linear between these two points, as also expected. This behavior is consistent with a strong microwave shielding effect present for thin metallic films when microwave power is incident on one film surface only [1,15].

Also, the shape of $Z_r(H)$ dependence as a function of the film thickness was checked. The results are shown in Fig. 2. The first observation from Fig. 2 is that two resonance peaks are present in Panels (b) and (c), although no surface pinning was assumed for either of the film surfaces. This is consistent with the calculation in [16] where it was explained as a consequence of a perfect microwave shielding effect. In the presence of the microwave eddy currents induced in the film by the incident microwave magnetic field of the microstrip line, the magnetization dynamics is driven by the sum of the spatially (quasi)-uniform Oersted field of the microwave current in the microstrip and the spatially anti-symmetric Oersted field of the eddy current in the film (see Fig. 21 in [1]). The uniform component is responsible for excitation of the fundamental mode of uniform precession (the right-hand peak in the panels of Fig. 2) and the anti-symmetric component for excitation of the 1st standing spin wave mode (1st SSW, the left-hand peak in the panels). The latter mode is characterized by an anti-symmetric distribution of dynamic magnetization across the film thickness.

The second observation from Fig. 2 is that an increase in $L$ leads to an upshift in the 1st standing spin wave mode peak. This behavior and the field positions of both peaks are consistent with Kittel Equation for the dipole-exchange modes [24]. This is one more confirmation of the validity of our theory. Noteworthy is the absence of the 2nd SSW peak in Fig. 2(b), although Kittel Equation predicts that this peak should be located at about 2000 Oe. Hence, the eddy current effect allows one to probe the 1st SSW mode only (unless a film possesses asymmetric surface magnetization pinning).

Next, we employ the developed numerical code in order to understand the dependence of the amplitude of the 1st SSW peak on the width of the microstrip $w$. In [16] it has been observed that the amplitude of this peak increases with an increase in $w$. Now, given the much higher computation speed of the present software, we are able to explore this effect in detail.

In order to perform this study, the calculations are repeated for a number of $w$ values. The obtained $Z_r(H)$ and S21($H$) traces are fitted with a complex function

$$F(H) = A_0 + \frac{A_1}{H - (H_1 + i\Delta H_1)} + \frac{A_2}{H - (H_2 + i\Delta H_2)} \quad . \tag{27}$$

Here $H_{1(2)}$ is the extracted resonance field for Mode 1 or 2 respectively, $\Delta H_{1(2)}$ is the linewidth of the respective peak. The quantities $A_0$, $A_1$ and $A_2$ are complex numbers; they are also extracted from the fits. In the following, we use quantities $|A_1/\Delta H_1|$ and $|A_2/\Delta H_2|$ to characterize the resonance peak heights.

The fits show that the shape of the $Z_r(H)$ dependence is in excellent agreement with Eq.(27). Hence, the complex function (27) is the "natural" dependence of $Z_r$ on $H$. This implies that the shape of the S21($H$) dependence may deviate from the one given by Eq.(27), because S21 depends on $Z_r$ in a complicated and nonlinear way (see Eq.(31) in [17]). Consequently, values of the parameters of Eq.(27) extracted from the fits may be different for $Z_r$ and S21 traces. This conclusion is confirmed by our numerical calculations.

Fig. 3 displays examples of traces obtained for different values of $w$. One sees that the 1SSW peak grows in amplitude with an increase in $w$. In [16] this was explained as a more pronounced microwave shielding effect for wider microstrips. The last panel of Fig. 3 shows the peak amplitude ratio $r = |A_2/\Delta H_2|/|A_1/\Delta H_1|$ as a function of $w$. (Here the index "1" denotes the fundamental mode and "2" the 1st SSW.). From this figure one sees that there is a strong dependence of the amplitude of the 1st SSW mode on the microstrip width for $w$ values below 300 micron. For larger $w$ values the dependence saturates and the effect becomes the same as in the case of normal incidence of a travelling plane electromagnetic wave on the surface of a ferromagnetic film [25,26]. One also notices that the traces are slightly different for $Z_r$ and S21. As discussed above, this is consequence of the fundamental difference in the shapes of $Z_r(H)$ and S21($H$) dependences.

Similar behavior is observed as a function of the thickness of the dielectric spacer $s$ (Fig. 4(e).) From the examples of the raw traces in Fig. 4(a-c), one sees that $r$ increases with an increase in $s$. This happens because lifting the film with respect to the stripline makes the microwave Oersted field of the current in the stripline more spatially uniform at the position $y$ where the film is located. The more uniform is the field, the more pronounced the microwave shielding is. Hence, the effect of lifting the film is analogous to increasing $w$ while keeping $s$ constant.

This conclusion implies that the $r(s)$ dependence should be significant for small $w$ values only. Indeed, the graphs in Fig. 4 were obtained for $w$=10 microns. For significantly wider striplines ($w$>300 microns) this dependence is practically vanishing.

From the raw traces in Fig. 4 one also notices that the increase in $s$ is accompanied by a decrease in the heights of both peaks. This behavior is caused by a decrease of the strength of coupling of the magnetization dynamics to the driving magnetic field [9, 17]).

The change in the coupling strength also leads to a dependence of the resonance peak linewidth on $s$. From Fig. 4(e) one sees that the linewidth decreases with a decrease in $s$. This is consistent with the effect of radiation losses [9]. Interestingly, the linewidth broadening due to the radiation losses is more significant for the 1st SSW than for the fundamental mode, with broadening being the same for both S21 and $Z_r$. This is actually in qualitative agreement with the experiment in [9], see Fig. 3 (c) in that paper. On the contrary, the fundamental-mode linewidth for small values of $s$ is noticeably larger for S21 than for $Z_r$. The latter fact is in agreement with the exchange-free model from [17].

Let us now look at the effect of surface magnetization pinning on magnetization dynamics. Fig. 5 displays a number of S21($H$) traces calculated for different values of the pinning constant. As before, we assume that the films have large conductivity of metals; this leads to important peculiarities of the stripline FMR responses, as we will show below.

Let us first discuss the effect of symmetric pinning – the situation when $d_p^x(y=0) = d_p^x(y=L)$. As seen from Fig. 5 (a,b,e,f), the main impact of the symmetric pinning is a shift of the resonance peaks upwards or downwards, depending on the sign of the pinning constant. This is in full agreement with Kittel Equation for spin wave resonance frequencies [24]. No noticeable difference in $r$ is seen for Panels (e) and (f) with respect to Panel (a).

On the contrary, asymmetric pinning Fig.5(c,d,g,h) has a strong effect on $r$. This is because the symmetry is now doubly broken – on top of the asymmetry of $h_x(y)$ originating from the single-side incidence of the microwave field, there is also asymmetry in the film's magnetic parameters in the direction of the film thickness. This makes the value of $r$ dependent on the film orientation with respect to the microstrip line.

This effect was theoretically found in [1] based on consideration of a simple model $w=\infty$. In that work, it was suggested that it might be useful for experimental determination of the particular film surface (from the two) at which magnetization pinning is present. Fig. 5 confirms this funding with a rigorous calculation for a realistic value of $w$ and hence the practical importance of this effect.

## IV. CONCLUSION

In this work, we constructed a new two-dimensional model for calculation of the stripline ferromagnetic resonance response of metallic ferromagnetic films. Our model works much more rapidly and is capable of taking exchange interaction and surface magnetisation pinning into account. The acceleration was achieved by analytically solving the initial system of equations describing the dynamics.

We also conducted a number of computations with a numerical model which followed from this theory. The numerical code enabled us to explore a large parameter space for the problem. Our computations confirmed that microwave shielding by eddy currents induced in a ferromagnetic film strongly affects its stripline FMR response if the film has large conductivity. The eddy currents are excited in the film because of the single-side incidence of the microwave field on the film in the geometry of Fig. 1. They lead to excitation of the 1st Standing Spin Wave Mode. The amplitude of the respective peak in the raw FMR traces depends on the width of the microstrip –with an increase in the width the microwave magnetic field incident on the film becomes more spatially uniform which leads to more efficient shielding by the eddy currents.

The developed model also allowed us to investigate radiation losses for the FMR modes. It was found that they are larger for the 1st Standing Spin Wave Mode than for the Fundamental one, in qualitative agreement with an earlier experiment [9].

Also the effect of the surface magnetization pinning on the response was explored in the framework of this more rigorous model. It confirmed a conclusion from a previous work that the stripline FMR method can be used to extract the degree of symmetry of pinning, and if the pinning turns to be asymmetric to identify which of film surfaces is characterized by larger (smaller) pinning.

Appendix

By substituting Eq. (14) and Eq. (15) into Eqs. (11-12), we obtain

$$\left(\omega(K^2 - q^2) + \omega_M kq\right) m_x - i\left[\alpha\omega_M q^4 - (\Omega_K + \omega_M)q^2 + \Omega_k K^2 + v^2 \omega_M\right] m_y = 0$$
$$i\left[\alpha\omega_M q^4 - (\Omega_K + \omega_M)q^2 + (\Omega_k + \omega_M)K^2\right] m_x + \left(\omega(K^2 - q^2) - \omega_M kq\right) m_y = 0$$
(A1)

where $v^2 = i\omega\sigma\mu_0$, $\Omega_k = \omega_H + \omega_M \alpha k^2$, and $\Omega_K = \Omega_k + \omega_M \alpha K^2$.

Equating the determinant of the matrix of this system of equations to zero results in the characteristic equation (16). The coefficients of this equation are as follows:

$a = \alpha\omega_M$,

$b = \alpha\omega_M \left[2\Omega_K + \omega_M \alpha(k^2 + v^2) + \omega_M\right]$,

$c = \omega_M^2 \alpha^2 k^2 (3k^2 - 2v^2) + 2\alpha\omega_M \left[(k^2 + v^2)(\omega_M + \omega_H) + k^2 \omega_H\right] + \omega_H(\omega_M + \omega_H) - \omega^2$,

$d_0 = -\omega_M^2 \alpha^2 k^2 (k^2 + v^2) - \omega_M \alpha\, k^2 \left[2(k^2 + v^2)(\omega_M + \omega_H) - k^2 \omega_M\right] - v^2 \left[\omega_H(\omega_M + \omega_H) - \omega^2\right]$. (A2)

The three roots of the cubic equation (16) are as follows.

$Q_1 = D - R - C$

$$Q_2 = Q_3^* = \frac{R}{2} - B - \frac{D}{2} - \frac{i\sqrt{3}(R+D)}{2},$$
(A3)

where $B = \dfrac{b}{3a}$, $C = \dfrac{c}{3a}$, $R = \dfrac{C - B^2}{D}$, and $D = \left\{D_1 - B^3 - d/(2a) + 3BC/2\right\}^{1/3}$, and

$D_1 = \left[C^3 + \dfrac{B^3 d}{a} - \dfrac{3B^2 C^2}{4} - \dfrac{Bcd}{2a^2} + \dfrac{d^2}{4a^2}\right]^{1/2}$. Once we have obtained the Q-values, we are able to express $m_y$ in terms of $m_x$ with the help of Eq.(A1):

$m_y = C_{my} m_x$, (A4-1)

where $C_{my} = \dfrac{-i\left[\alpha\omega_M q^4 - (\Omega_K + \omega_M)q^2 + (\Omega_k + \omega_M)K^2\right]}{\omega(K^2 - q^2) - \omega_M kq}$. Also, the following relations between the other field components follow from Eqs.(14)-(15):

$$h_x = C_{hx} m_x, \qquad (A4\text{-}2)$$

where $C_{hx} = -(K^2 + qkC_{my})/(K^2 - q^2)$, and

$$h_y = C_{hy} m_x, \qquad (A4\text{-}3)$$

where $C_{hy} = -C_{my} + ik(C_{hx} + 1)/q$.

The system of equations which follows from the boundary value problem is given by:

$$\sum_{i=1}^{6}(q_i + d_p^x)M_x^i = 0,$$

$$\sum_{i=1}^{6}(q_i - d_p^x)M_x^i \exp(q_i L) = 0,$$

$$\sum_{i=1}^{6}(q_i + d_p^y)C_{my}(q_i)M_x^i = 0,$$

$$\sum_{i=1}^{6}(q_i - d_p^y)C_{my}(q_i)M_x^i \exp(q_i L) = 0, \qquad (A5)$$

$$\sum_{i=1}^{6}\left\{\left[C_{my}(q_i) + C_{hy}(q_i)\right]\coth\left[|k|(d+s)\right] + i\,\mathrm{sgn}(k)C_{hx}(q_i)\right\} M_x^i = \frac{i\,\mathrm{sgn}(k)\sinh(|k|d)}{\sinh\left[|k|(d+s)\right]},$$

$$\sum_{i=1}^{6}\left[C_{my}(q_i) + C_{hy}(q_i) + i\,\mathrm{sgn}(k)C_{hx}(q_i)\right]\exp[q_i L]M_x^i = 0.$$

In this work this system is solved numerically to obtain the six coefficients $M_i^x$.

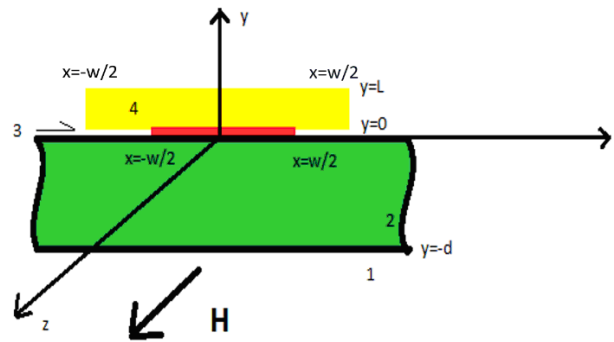

Figure 1. Sketch of the modeled geometry. (1) The ground plane of the microstrip line. (2) Substrate of the microstrip line of thickness $d$. (3) Infinitely thin strip of width $w$ carrying a microwave current in the direction $z$. (4) Ferromagnetic film of thickness $L$ in the direction y and of width $w$ in the direction x. The static magnetic field $H$ is applied along z.

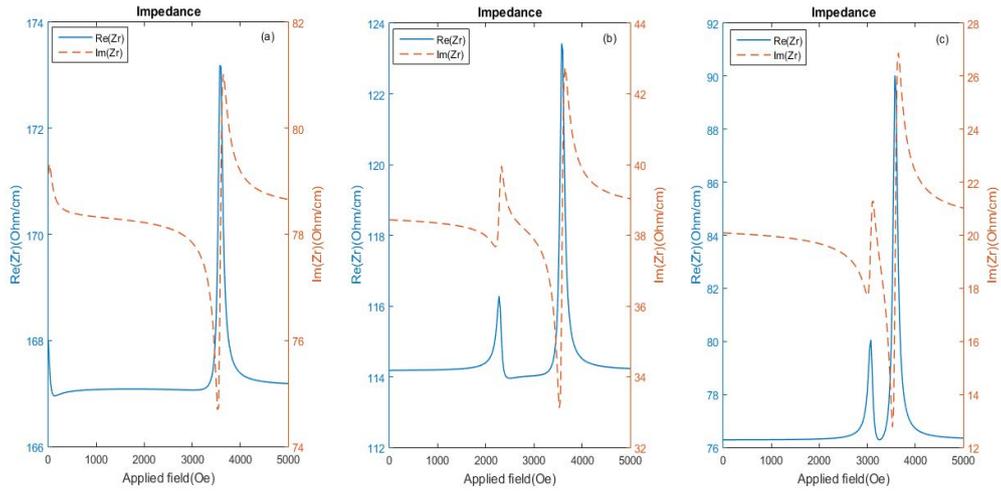

Figure 2. Calculated FMR traces for different ferromagnetic film thicknesses $L$. (a) $L$=30nm; (b) $L$=50nm; (c) $L$=80nm. Parameters of calculation: width of microstrip line $w$=350μm; thickness of the line substrate d=300μm; spacer thickness $s$=1μm; microwave frequency is 20GHz; saturation magnetization $4\pi M_s$=10000G; conductivity of the film $\sigma$=4.5·10$^6$S/m; magnetic loss parameter for the ferromagnetic film: $\Delta H$=53.017 Oe (Gilbert damping constant $\alpha_G$=0.008), exchange constant $A$=0.8·10$^{-6}$ erg/cm. Unpinned surface spins ($d_p^x = d_p^y = 0$) are assumed for both film surfaces.

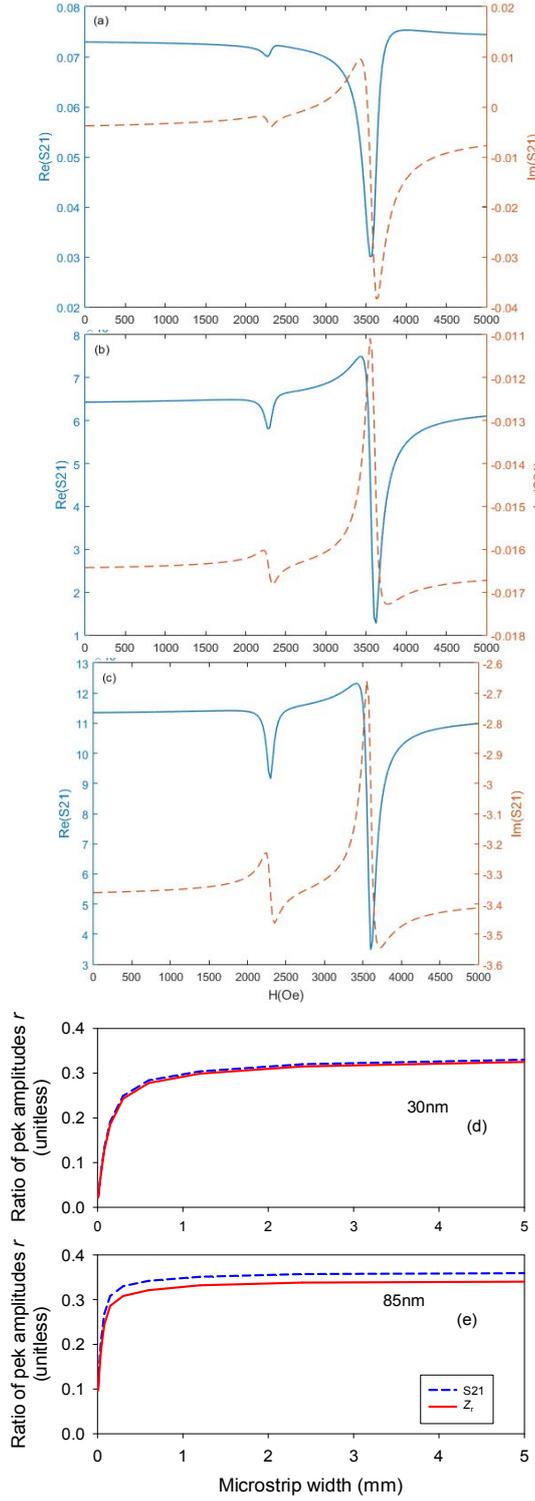

Fig. 3. Panels (a) to (c): FMR traces for different microstrip line widths $w$=10, 100 and 350μm respectively. Other parameters are the same as for Fig.2(a). (d) and (e): dependence of the ratio of the peak amplitudes on $w$ for two film thicknesses $L$=30nm (d) and $L$=85nm (e). Parameters of calculation for (d) and (e): $L$=85 micron, $4\pi M_s$=10000G, $\sigma$=4.5·10⁶S/m, exchange constant $A$=1.3·10$^{-6}$erg/cm, Gilbert damping constant 0.008, $w$=0.1 mm, $s$=10micron, spacer dielectric permittivity: 1; microstrip substrate thickness $d$=0.1mm and permittivity 10. Length of the film in the direction along the microstrip is 5 mm. Frequency is 22 GHz.

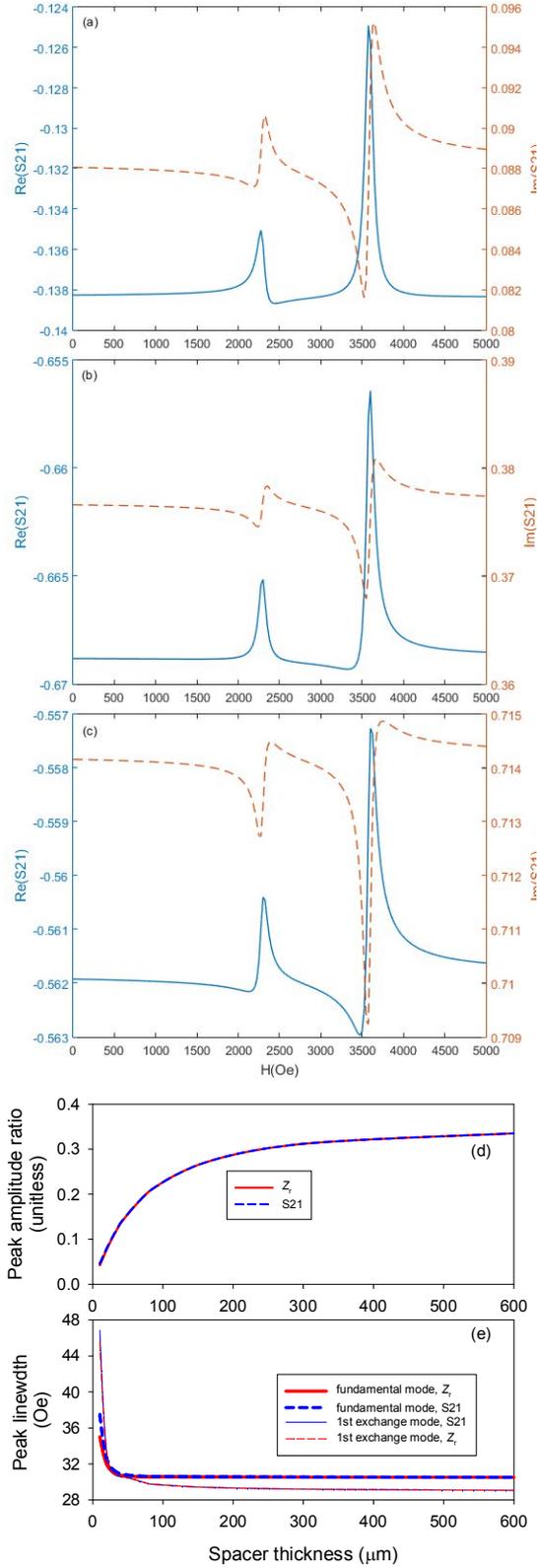

Fig. 4. Panels (a) to (c) are plotted for spacers of thickness $s$=10, 100 and 200 microns respectively. The microstrip line width is 10 micron and other parameters are same as in Fig.2(a). (d): dependence of the peak amplitude ratio on $s$. (e) Peak linewidth dependence on $s$. Parameters of calculation for (d) and (e): $L$=85 nm, $w$=10μm, frequency is 10 GHz. The remainder of parameters is the same as for Fig. 3(d-e).

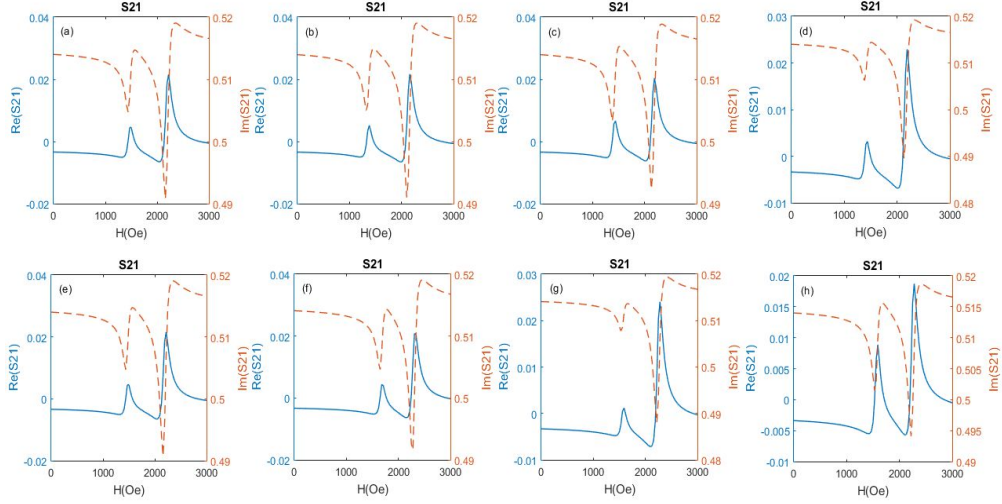

Figure 5. Examples of FMR traces for different surface magnetization pinning conditions. (a) and (e): magnetisation is unpinned $d_p^y = 0$ at both film surfaces. (b) and (f): it is equally pinned at both surfaces. The pinning constant $d_p^y = 10^8 \, \text{m}^{-1}$ for (b) and $d_p^y = -10^8 \, \text{m}^{-1}$ for (f). (c) and (g): single-sided pinning. $d_p^y = 0$ for $y=L$ and $d_p^y = 10^8 \, \text{m}^{-1}$ (c) or $d_p^y = -10^8 \, \text{m}^{-1}$ (g) for $y=0$. (d) and (h): the same, but now $d_p^y = 0$ for $y=0$, and $d_p^y = 10^8 \, \text{m}^{-1}$ (d) or $d_p^y = -10^8 \, \text{m}^{-1}$ (h) for $y=L$. $d_p^x = 0$ for all panels. Width of the microstrip line $w$=350 μm, thickness of the film $L$=50nm; thickness of the spacer $s$=0; microwave frequency is 19 GHz; saturation magnetization $4\pi M_s$=17900G; conductivity of the film $\sigma$=1.8·10$^7$S/m; Gilbert damping constant is 0.008. Note that that (a) and (e) show the same plot; the plot is repeated to facilitate its comparison with the other plots.